\RequirePackage{ifpdf}
\ifpdf 
\documentclass[pdftex]{sigma}
\else
\documentclass{sigma}
\fi

\usepackage[ukrainian,english]{babel}
\usepackage[cp1251]{inputenc}

\newcommand{\gmn}{g_{MN}}
\newcommand{\gmu}{g_{\mu \nu}}
\newcommand{\ckl}{C^{KLMN}}
\newcommand{\rkl}{R^{KLMN}}

\begin{document}

\renewcommand{\PaperNumber}{091}

\FirstPageHeading

\renewcommand{\thefootnote}{$\star$}

\ShortArticleName{Dimensional Reduction of Conformal Tensors and Einstein--Weyl Spaces}

\ArticleName{Dimensional Reduction of Conformal Tensors\\ and Einstein--Weyl Spaces\footnote{This paper is a contribution to the Proceedings of
the Seventh International Conference ``Symmetry in Nonlinear
Mathematical Physics'' (June 24--30, 2007, Kyiv, Ukraine). The
full collection is available at
\href{http://www.emis.de/journals/SIGMA/symmetry2007.html}{http://www.emis.de/journals/SIGMA/symmetry2007.html}}}

\Author{Roman  JACKIW}

\AuthorNameForHeading{R.~Jackiw}

\Address{Department of Physics, Massachusetts Institute of Technology, Cambridge, MA 02139, USA}
\Email{\href{mailto:jackiw@mit.edu}{jackiw@mit.edu}}

\ArticleDates{Received August 29, 2007, in f\/inal form September 16,
2007; Published online September 21, 2007}

\Abstract{Conformal Weyl and Cotton tensors are dimensionally reduced by a Kaluza--Klein procedure. Explicit formulas are given for reducing from four and three dimensions to three and two dimensions, respectively. When the higher dimensional conformal tensor vanishes because the space is conformallly f\/lat, the lower-dimensional Kaluza--Klein functions satisfy equations that coincide with the Einstein--Weyl equations in three dimensions and kink equations in two dimensions.}


\Classification{81T40}

\section{Introduction}

In this paper, we shall examine a mathematical problem, which does not possess any evident physical signif\/icance, but nevertheless leads to interesting equations that lie in various areas of mathematical physics. The problem that we address concerns the dimensional reduction of geometric conformal tensors (def\/ined below) from $n$ to $n-1$ dimensions. More specif\/ically, for an $n$-dimensional conformal tensor, constructed from a metric tensor $\gmn, \{ M, N\}: 0, 1, \dots, n-1$, the metric tensor is parameterized in the Kaluza--Klein fashion
\begin{gather}
\gmn = e^{2\sigma}
\left(
\begin{array}{rr}
\gmu - a_\mu a_\nu & -a_\mu \\
-a_\nu & -1
\end{array}
\right),\qquad
\{\mu, \nu\}: 0,1,\dots, n-2,\label{rjeq1}
\end{gather}
corresponding to the $n $ and $n-1$ line elements
\begin{gather*}
ds^2_{(n)} = \gmn \, dx^M\, dx^N = e^{2\sigma} \big [ ds^2_{n-1} - (a_\mu \, dx^\mu + dx^{n-1})^2\big],\\
ds^2_{(n-1)} = \gmu\, dx^\mu \, dx^\nu. \nonumber
\end{gather*}
The metric functions are taken to be independent of the ``last'' coordinate $x^{n-1}$. Consequently the $n$-dimensional conformal tensor reduces to an $n-1$ dimensional conformal tensor, plus other geometrical entities appropriate to $n-1$ dimensions. An $n$-dimensional dif\/feomorphism, $ \delta x^N = -f^N (x)$, where the transformation function is independent of the last coordinate, acts on the quantities in~\eqref{rjeq1} in such a way that $\{\gmu, a_\mu, \sigma\}$ transform as $n-1$ dimensional tensors, vector and scalar, respectively, and also $a_\mu$ experiences an Abelian gauge transformation.

The ef\/fect of such a dimensional reduction on the $n$-dimensional Riemann tensor is known: it is described by the classic geometric Gauss--Codazzi equations (see for example \cite{rjbib1}). Here we present analogous results for conformal tensors, in various dimensions.

We begin by describing conformal tensors.

\section{Conformal tensors}

In dimensions greater than three, the conformal tensor is the Weyl tensor $\ckl$ related to the Riemann tensor by
\begin{gather}
\ckl \equiv \rkl  -\frac{1}{n-2}\big(g^{KN} \, S^{ML} - g^{KN} \, S^{ML} - g^{LM}\, S^{NK} + g^{LN}\, S^{MK}\big),\label{rjeq3}\\
 S^{NL} \equiv R^{NL} - \frac{1}{2(n-1)}\, g^{NL}\, R. \nonumber
\end{gather}
$R^{NL}$ and $R$ are the Ricci quantities, while $\ckl$ has the following properties
\begin{description}\itemsep=0pt
\item{(a)} It is covariant against conformal redef\/inition of the metric \label{itema}
\[
\gmn (x) \to \lambda (x) \gmn (x).
\]
\item{(b)} It vanishes if and only if the space is conformally f\/lat: $\gmn$ is dif\/feomorphic to $\lambda \eta_{MN}$ where $\eta_{MN}$ is f\/lat. \label{itemb}
\item{(c)} It possesses the symmetries of the Riemann tensor and also is traceless in each index pair.\label{itemc}
\end{description}

Evidently, according to properties (a) and (b), the Weyl tensor acts as a template for conformal f\/latness: by evaluating it on a specif\/ic metric tensor one can learn whether the space-time is conformally f\/lat.

In three dimensions the Weyl tensor vanishes identically, and the Riemann tensor is given by the last term in \eqref{rjeq3} (at $n = 3$). But not all 3-dimensional space-times are conformally f\/lat. So there is needed a replacement for the Weyl tensor, which would act as a template for conformal f\/latness.

The Cotton tensor
\begin{gather*}
C^{KL} = \frac{1}{2\sqrt{g}} \big(\varepsilon^{KMN}\, D_M\, R^L_N + \varepsilon^{LMN} \, D_M \, R^K_N\big)
\end{gather*}
serves that role, as it possesses conformal template properties (a) and (b). $C^{KL}$ is symmetric in its indices, and like the Weyl tensor, it is traceless. The Weyl tensor is not obtained by varying a scalar action. But the Cotton tensor enjoys this property, since it is the variation of the gravitational Chern--Simons term
\begin{gather*}
CS (\Gamma) \equiv \frac{1}{4 \pi^2} \int d^3 x \, \varepsilon^{KLM} \left(\frac{1}{2}\, \Gamma^R_{KS}\, \partial_L \, \Gamma^S_{MR} + \frac{1}{3}\, \Gamma^R_{KS}\, \Gamma^S_{LT}\, \Gamma^T_{MK}\right),\\ 
\delta CS(\Gamma) = -\frac{1}{4 \pi^2} \int d^3 x \, \sqrt{g}\ C^{K L}\, \delta g_{KL}. 
\end{gather*}
Because $CS(\Gamma)$ is a scalar, $C^{KL}$ is covariantly  conserved. 

Finally in two dimensions, all spaces are (locally) conformally f\/lat, so there is no need for a~conformal tensor, and indeed none exists.

\section{Dimensional reductions}

\quad We have derived the relevant formulas for the general $n\to  n-1$ reduction. They are complicated and will not be recorded here, since they appear in the published research paper \cite{Grumiller:2006ww}. We remark on two general features
\begin{description}\itemsep=0pt
\item{(a)} The reductions of the Weyl and Cotton tensors do not depend on the conformal factor $e^{2\sigma}$ in \eqref{rjeq1}, because the tensors are conformally invariant.

\item{(b)} The dependence of the tensors on the vector $a_\mu$ is mostly through the ``gauge invariant'' combination.
\begin{gather}
f_{\mu \nu} = \partial_\mu\, a_\nu - \partial_\nu \, a_\mu.
\label{rjeq8}
\end{gather}
\end{description}

We present explicit formulas for the $4\to 3$ and $3\to 2$ reductions. These are especially interesting owing to the absence of a Weyl tensor in three and two dimensions, and the presence of the Cotton tensor in three dimensions.

\subsection[$4\to 3$ reduction]{$\boldsymbol{4\to 3}$ reduction}

The 4-dimensional Weyl tensor $C^{MNKL}$  with indices in the 3-dimensional range ($\mu, \nu, \lambda, \tau$) reads
\begin{gather}
C^{\mu\nu\lambda\tau} = g^{\mu\lambda}\, c^{\tau\nu} - g^{\mu\tau} c^{\lambda\nu} - g^{\nu\lambda} \, c^{\tau\mu} + g^{\nu\tau}\, c^{\lambda\mu},
\label{rjeq9}
\end{gather}
where
\begin{gather}
c^{\mu\nu} \equiv \frac{1}{2} \left(r^{\mu\nu} - \frac{1}{3}\, g^{\mu\nu}\, r - f^\mu f^\nu + \frac{1}{3} \, g^{\mu\nu}\, f^2\right).
\label{rjeq10}
\end{gather}
Here $r^{\mu\nu}$ and $r$ are  3-dimensional Ricci entities, constructed from the 3-dimensional metric tensor~$g_{\mu\nu}$, while $f^\mu$ is the dual ``f\/ield'' strength, constructed from the vector potential $a_\mu$ (see~\eqref{rjeq8})
\begin{gather*}
f^\mu \equiv \frac{\epsilon^{\mu\alpha\beta}}{2\sqrt{g}}\, f_{\alpha \beta} = \frac{\varepsilon^{\mu\alpha\beta}}{\sqrt{g}}\, \partial_\alpha\, a_\beta.
\end{gather*}

There  remains one more component of the 4-dimensional Weyl tensor that must be specif\/ied
\[
C^{-\lambda\mu\nu} =  \frac{\epsilon^{\mu\nu\tau}}{2\sqrt{g}}  \big(d^\lambda f_\tau + d_\tau f^\lambda\big) - a_\tau\, C^{\tau\lambda\mu\nu}.
\]
Here $d^\lambda$ is the 3-dimensional covariant derivative, and ``$-$'' refers to the ``last'' ($M=3$) va\-lued index of the 4-dimensional Weyl tensor. Other components of $C^{MNKL}$ are f\/ixed by its tracelessness~\cite{Grumiller:2006ww}.

\subsection[$3\to 2$ reduction]{$\boldsymbol{3\to 2}$ reduction}

The 3-dimensional Cotton tensor $C^{MN}$, with its indices in the 2 dimensional range, becomes
\begin{gather}
C_{\mu\nu} = \gmu \left(d^2 f - f^3 -\frac{1}{2}\, r f\right) - d_\mu d_\nu\, f.
\label{rjeq12}
\end{gather}
Now $r$ is the 2-dimensional Ricci scalar constructed from the 2-dimensional metric tensor $\gmu$; $d_\mu$ is the appropriate 2-dimensional covariant derivative; $f$ is the dual to $f_{\mu\nu}$ -- a scalar in two dimensions
\begin{gather}
f \equiv \frac{\varepsilon^{\mu\nu}}{2\sqrt{-g}}\, f_{\mu\nu} = \frac{\varepsilon^{\mu\nu}}{\sqrt{-g}}\, \partial_\mu a_\nu.
\label{rjeq13}
\end{gather}
The further component of the Cotton tensor is given by
\begin{gather*}
C^{-\mu} = \frac{1}{2\sqrt{-g}}\, \varepsilon^{\mu\nu}\, \partial_{\nu} \, (r + 3f^2) - a_\nu\, C^{\mu\nu}
\end{gather*}
with ``$-$'' denoting the ``last'' ($M=2$) valued index. The remaining component of $C^{MN}$ is f\/ixed by its tracelessness~\cite{Guralnik:2003we}.

\section[Embedding in a conformally flat space]{Embedding in a conformally f\/lat space}

If we demand that the higher dimensional conformal tensor vanishes, the reduced formulas become equations that determine the lower-dimensional metric tensor and vector f\/ield (actually only its ``gauge''-invariant curl enters). The lower dimensional geometry is therefore embedded in a conformally f\/lat space of one higher dimension.

\subsection[$4\to 3$ embedding]{$\boldsymbol{4\to 3}$ embedding}

 When the 4-dimensional Weyl tensor vanishes \eqref{rjeq9} and \eqref{rjeq10} imply the 3-dimensional traceless equation
\begin{gather}
C^{\mu\nu} = 0 \quad \Rightarrow \quad r^{\mu\nu} -\frac{1}{3}\, g^{\mu\nu}\, r = f^\mu f^\nu - \frac{1}{3}\, g^{\mu\nu} \, f^\alpha f_\alpha .
\label{rjeq15}
\end{gather}
While \eqref{rjeq9} and \eqref{rjeq12} require
\begin{gather}
d_\mu f_\nu + d_\nu f_\mu = 0,
\label{rjeq16}
\end{gather}
{\it i.e.} $f^\mu$ is a Killing vector of the 3-geometry.
Equations \eqref{rjeq15} and \eqref{rjeq16} have the consequence (by dif\/ferention of \eqref{rjeq15} and use of \eqref{rjeq16})
\begin{gather*}
r = -5 f^2 + c,
\end{gather*}
where $c$ is a constant. Also one readily shows that the quantity
\begin{gather*}
F^\mu \equiv \frac{\varepsilon^{\mu\nu\lambda}}{\sqrt{g}}\ d_\nu f_\lambda
\end{gather*}
is an additional Killing vector of the 3-geometry~-- we call it the ``dual'' Killing vector \cite{Grumiller:2006ww}.

We present explicit solutions to \eqref{rjeq15} and \eqref{rjeq16} that are static and circularly symmetric.
With such an Ansatz, two solutions are found \cite{Grumiller:2006ww}
\begin{alignat}{3}
&(a)\quad &&  ds^2_3 = v (\rho) d t^2 - \frac{4/a}{1-\rho^2/a}\, \frac{d\rho^2}{v(\rho)} - \rho^2 d \theta^2,\label{rjeq19}&\\
&&& v(\rho) \equiv A + B \sqrt{1-\rho^2/a}.\nonumber&
\end{alignat}
$f^\mu$ is the time-like Killing vector for \eqref{rjeq19}
\begin{gather}
f^\mu : \  \{f^t = 1, f^\rho = 0, f^\theta = 0\},
\label{rjeq20}
\end{gather}
while the dual Killing vector is spacelike
\begin{gather}
F^\mu :  \ \{F^t = 0, F^\rho = 0, F^\theta = 1\}.
\label{rjeq21}
\end{gather}
Note  $|a|$ may be eliminated from \eqref{rjeq19} by rescaling $\rho$ and $\theta$.
\begin{alignat}{3}
& (b)\quad  && ds^2_3 = w (\rho) d t^2 - \frac{1}{w (\rho)}\, d\rho^2 - \rho^2 (d \theta)^2, \label{rjeq22}&& \\
&&& w (\rho) \equiv \frac{1}{4}\ \rho^4 + A \rho^2 + B. \label{rjeq23}&
\end{alignat}
Now $f^\mu$ is the space-like Killing vector,
\begin{gather}
f^\mu = \{f^t = 0, f^\rho = 0, f^\theta = 1\}
\label{rjeq24}
\end{gather}
while dual Killing vector is time-like
\begin{gather}
F^\mu : \ \{F^t =1, F^\rho=0, F^\theta =0\}.
\label{rjeq25}
\end{gather}

\subsection[$3\to 2$ embedding]{$\boldsymbol{3\to 2}$ embedding}

The vanishing of the 3-dimensional Cotton tensor requires, according to \eqref{rjeq12}, both the 2-di\-men\-sio\-nal traceless equation
\begin{gather}
\left(d_\mu d_\nu - \frac{g_{\mu\nu}}{2}\ d^2\right) f = 0
\label{rjeq26}
\end{gather}
and the trace condition
\begin{gather}
d^2 f -2 f^3 -r f = 0
\label{rjeq27}
\end{gather}
while \eqref{rjeq13} sets
\begin{gather}
r = - 3 f^2 +c
\label{rjeq28}
\end{gather}
where $c$ is a constant. Therefore \eqref{rjeq27} becomes \cite{Guralnik:2003we}
\begin{gather}
d^2 f + f^3 - c f = 0.
\label{rjeq29}
\end{gather}
Since, unlike the Weyl tensor, the Cotton tensor is the variation of an action -- the gravitational Chern--Simons action $CS (\Gamma)$ -- equations \eqref{rjeq26}--\eqref{rjeq29} arise by varying the dimensionally reduced $CS(\Gamma)$, which reads
\begin{gather}
CS = -\frac{1}{8\pi^2}  \int d^2 x\,  \sqrt{-g} \, (f r + f^3).
\label{rjeq30}
\end{gather}

The equations \eqref{rjeq26}--\eqref{rjeq29} can be solved for arbitrary values of $c$: positive, negative, zero \cite{Guralnik:2003we,rjbib4}. Especially interesting are the solutions for $c>0$, where the $f \leftrightarrow -f$ ref\/lection ``symmetry'' of the equations (not of the action \eqref{rjeq30}) is spontaneously broken by the solution $f = \pm \sqrt{c}$, $r= 2 c$. In further analogy with familiar f\/ield theoretical behavior, there also exists a solution which interpolates between the $\pm \sqrt{c}$ ``vacua''
\begin{gather*}
f = \sqrt{c} \tanh \ \frac{\sqrt{c} x}{2},\qquad
r= 2 c + \frac{3c}{\cosh^2 \frac{\sqrt{c}}{2}x}.
\end{gather*}

It is amusing to recall the above mentioned f\/ield theoretic analog. In 2-dimensional Minkowski space-time a scalar f\/ield  $\Phi$ can satisfy the equation
\begin{gather*}
\Box \Phi  + \Phi^3 - c \Phi = 0, \qquad c>0
\end{gather*}
which possesses the $\Phi \leftrightarrow - \Phi$ symmetry breaking solution $\Phi = \pm \sqrt{c}$. As is well known the equation also admits the kink solution which interpolates between the  two $ \pm \sqrt{c}$ ``vacua''
\[
\psi_{\mbox{kink}} = \sqrt{c}\, \tanh \sqrt{\frac{c}{2}}\, x.
\]

The analogy is perfect, but there is a normalization discrepancy. The curved-space kink possess a spatial coordinate scaled by 2, while in the f\/lat-space kink the scaling is $\sqrt{2}$. This is the only ef\/fect of the non-trivial geometry.

\section{Other work}

In \cite{rjbib4,rjbib5} we list papers that rely to some extent on the results presented here, specif\/ically employing formulas arising in the $3\to 2$ reduction.

In the discussion following the lecture, M.~Eastwood  observed that the formulas relevant to the $4\to 3$ reduction coincide with equations that arise in the 3-dimensional Einstein--Weyl conformal theory \cite{rjbib6}. This is an interesting connection, which we now elaborate.

Einstein--Weyl theory (in any dimension) is equipped with a metric tensor $g_{\mu\nu}$ and an additional vector $W_\omega$ -- the ``Weyl potential'' -- which arises when the covariant ``Weyl derivative''~$\triangle^W_\omega$, involving the torsion-less ``Weyl connection'' $W^\lambda_{\mu\nu}$, acts on $\gmu$ and preserves its conformal class~\cite{rjbib1,rjbib7}
\begin{gather}
\triangle^W_\omega \gmu \equiv \partial_\omega \, \gmu - W^\lambda_{\omega\mu} \, g_{\lambda\nu} - W^\lambda_{\omega\nu}\, g_{\mu\lambda} = 2 W_\omega\, g_{\mu\nu}.
\label{rjeq33}
\end{gather}
The Weyl connection can be constructed from the conventional Christof\/fel connection $\Gamma^\lambda_{\mu\nu}$, supplemented by an $W_\omega$-dependent expression
\begin{gather}
W^\lambda_{\mu\nu} = \Gamma^\lambda_{\mu\nu} + W^\lambda\, \gmu - W_\mu\, \delta^\lambda_\nu - W_\nu\, \delta^\lambda_\mu.\label{rjeq34}
\end{gather}

A curvature tensor is determined as usual by
\begin{gather}
[\triangle^W_\mu, \triangle^W_\nu] V_\alpha = - {^W\!\!R^\beta}_{\alpha\mu\nu} \, V_\beta,
\label{rjeq35}
\end{gather}
whose traces def\/ine ``Ricci'' quantities
\begin{gather*}
{^W\!\! R_{\mu\nu}} = {^W\!\! R^\alpha}_{\mu\alpha\nu},
\qquad
{^W\!\! R} =  {^W\!\! R^\mu}_{\mu}.
\end{gather*}

The Einstein--Weyl equation then requires that ${^W\!\! R}_{(\mu\nu)}$, the symmetric part of the ``Ricci'' tensor (generically ${^W\!\! R}_{(\mu\nu)}$ is not symmetric), be in the same conformal class as the metric tensor,
\begin{gather*}
{^W \!\! R}_{(\mu\nu)} = \lambda \, \gmu
\end{gather*}
or equivalently in three dimensions
\begin{gather}
{^W\!\! R}_{(\mu\nu)}  - \frac{\gmu}{3}\, {^W R} = 0.
\label{rjeq39}
\end{gather}
From \eqref{rjeq34} and \eqref{rjeq35} ${^W R}_{(\mu\nu)}$ can be expressed in terms of the usual Ricci tensor, supplemented by $W_\omega$-dependent terms
\begin{gather}
{^W\!\! R}_{(\mu\nu)} = R_{\mu\nu} + D _{(\mu} W_{\nu)} + W_\mu W_\nu + \gmu  (D_\lambda \, W^\lambda - W_\lambda W^\lambda).
\label{rjeq40}
\end{gather}
Thus the Einstein--Weyl equation \eqref{rjeq39} requires the vanishing of a trace free quantity.
\begin{gather}
R_{\mu\nu} -\frac{1}{3}\, \gmu\, R + W_\mu W_\nu - \frac{\gmu}{3}\, W^\lambda W_\lambda
+ D _{(\mu}W_{\nu)} -\frac{1}{3}\, \gmu \, D^\lambda W_\lambda =0.
\label{rjeq41}
\end{gather}
(In \eqref{rjeq40} and \eqref{rjeq41} $D_\omega$ is the covariant derivative constructed with the Christof\/fel connection~$\Gamma^\alpha_{\mu\nu}$.)

The equations \eqref{rjeq33} and \eqref{rjeq40} are preserved under conformal transformations: the metric tensor is rescaled and the Weyl potential undergoes a gauge transformation.
\begin{gather*}
\gmu \to e^{2\sigma}\, \gmu, \qquad W_\mu \to W_\mu + \partial_\mu \, \sigma.
\end{gather*}
This gauge freedom is f\/ixed by choosing the ``Gauduchon'' gauge $D^\mu W_\mu = 0$, and one can further show that for positive def\/inite metrics $D _{(\mu}\!W_{\!\nu)}$ vanishes
\begin{gather*}
D _{(\mu} W_{ \nu)} = 0.
\end{gather*}
This leaves from \eqref{rjeq41} the gauge f\/ixed, 3-dimensional Einstein--Weyl equation
\begin{gather*}
R_{\mu\nu} -\frac{1}{3}\, \gmu\, R + W_\mu W_\nu - \frac{\gmu}{3}\, W^\lambda W_\lambda = 0.
\end{gather*}

Comparison with \eqref{rjeq15} and \eqref{rjeq16} shows that our 3-dimensional equation for the vanishing of the 4-dimensional Weyl tensor coincide with the gauge-f\/ixed Einstein--Weyl equation, apart from a~relative sign between the curvature quantities and the Weyl quantities. This sign discrepancy arises because we performed our reduction on a space with indef\/inite signature. When the reduction is performed with positive metric, the signs coincide.

This relation to Einstein--Weyl theory puts our work into contact with a wide mathematical literature where solutions other than our \eqref{rjeq19}--\eqref{rjeq25} are derived.

\subsection*{Acknowledgements}
This work was supported in part by funds provided by the U.S. Department of Energy under cooperative research agreement
\#DF-FC02-94ER40818.

\bigskip



{\it  {\selectlanguage{ukrainian} Дуже мені приємно сказати слова на закінчення конференції, але по Англійскі.}

 It was very useful and pleasant to participate in this
conference. The pleasure came from seeing how active and vibrant
physics, mathematics and mathematical physics are in Eastern
Europe, specifically Ukraine. The usefulness for me was in the
informative comments that I~received about my presentation. They
will certainly inform my further research.

I am sure that all other participants have similarly
benefited from and enjoyed the meeting, which was efficiently run
by the local organizers. All of  us thank them for their warm
Ukrainian hospitality and we look forward to returning in the
 future to charming Kiev.}

\pdfbookmark[1]{References}{ref}
\LastPageEnding

\end{document}